# Aptamer-protein interaction prediction model based on transformer

Zhichao Yan, Yue Kang, Buyong Ma

## Abstract

Aptamers are single-stranded DNA/RNAs or short peptides with unique tertiary structures that selectively bind to specific targets. They have great potential in the detection and medical fields.

Here, we present SelfTrans-Ensemble, a deep learning model that integrates sequence information models and structural information models to extract multi-scale features for predicting aptamer-protein interactions (APIs). The model employs two pre-trained models, ProtBert and RNA-FM, to encode protein and aptamer sequences, along with features generated from primary sequence and secondary structural information. To address the data imbalance in the aptamer dataset imbalance, we incorporated short RNA-protein interaction data in the training set. This resulted in a training accuracy of 98.9% and a test accuracy of 88.0%, demonstrating the model's effectiveness in accurately predicting APIs. Additionally, analysis using molecular simulation indicated that SelfTrans-Ensemble is sensitive to aptamer sequence mutations. We anticipate that SelfTrans-Ensemble can offer a more efficient and rapid process for aptamer screening.

## Introduction

Aptamers, a class of biomolecules introduced in 1990[1], has drawn significant attention in light of the emerging prominence of nucleic acid-based therapeutics. Aptamers are single-stranded oligonucleotides or short peptides characterized by a distinctive three-dimensional architecture comprising of 20 to 100 nucleotides (nt). They exhibit high affinity and specificity towards target molecules. The development of aptamers is primarily through the in vitro screening process known as Systematic Evolution of Ligands by Exponential Enrichment (SELEX)[2]. This method starts with a randomized oligonucleotide library with $10^{12}$ to $10^{15}$ individual oligonucleotides. Followed by iterative cycles of isolation and enrichment steps to generate highly specific and high-affinity aptamers tailored to the desired targets.

Aptamers are often referred to as chemical antibodies due to their ability to bind to specific molecules. They offer several advantages over traditional antibodies. Aptamers have a wider range of target molecules, including metal ions[3], small molecules[4], proteins[5], cells[6], and even organs and tissues[7, 8]. Aptamers can be easily produced in large quantities with low lot-to-lot variability and possess simple chemical structures, allowing them to be easily chemically modified by the addition of various functional groups. With low molecular weight, aptamers exhibit excellent tissue permeability and low immunogenicity. Therefore, as novel recognition molecules, aptamers are expected to have great potential in detection and therapy.

The SELEX technique is an empirical experimental method. Aptamers obtained by this method are often time-consuming to produce and may have low affinity. So far, many variant SELEX techniques[9, 10] and post-SELEX optimization techniques[11, 12] have been developed to speed up the screening time and improve the affinity of aptamers. However, there is a need for more efficient methods to expedite the screening process and enhance the affinity of aptamers. With the development of computer technology and hardware, artificial intelligence algorithms have demonstrated excellent performance in the field of nucleic acids. Supervised machine learning can predict unlabeled data using labeled data. RaptRanker[13] selects candidate aptamer sequences based on sequence frequency using local sequence motifs and structural information, which can accelerate SELEX screening time. Frances et al.[14] utilized the CountVectorizer method to extract information from the aptamer sequences followed with Support Vector Machines(SVM) to classify aptamer and non-aptamer sequences in the SELEX screening process. Aptamer-protein interactions (APIs) can also be predicted by this method. In 2020, Li et al. [15] integrated Adaboost and random forest (RF) algorithms to develop the model called PPAI to predict APIs. Features such as nucleotide composition, pseudo-nucleotide composition (PseKNC), and normalized Moreau-Broto autocorrelation coefficient were used to encode Aptamers, while proteins were encoded using amino acid composition (AAC), pseudo-amino acid composition (PseAAC), grouped amino acid composition, C/T/D composition, and sequence-order-coupling number. Similarly, in 2021, Emami et al. [16] developed a model for APIs prediction called Aptanet, using K-mers to encode aptamers and PseAAC to encode proteins. The classification prediction is trained through Multi-Layer Perceptron (MLP) classifiers. While these sequence attributes possess meaningful interpretations, they may not fully encompass the overall characteristics of aptamer and protein sequences. Natural Language Processing (NLP) techniques, on the other hand, excel at linking contextual relationships. Transformer-based deep learning models have become increasingly prevalent in recent years, particularly in the fields of structure prediction[17], protein-protein interactions[18], and RNA-protein interactions. BERT-RBP[19] has successfully predicted RNA-protein interactions using the encoder component of the transformer. AptaTrans[20] employed a transformer-based encoder for sequence embedding and mutual matrix computation between aptamer and protein monomers, facilitating APIs prediction.

In this study, we propose an integrated deep learning-based approach, SelfTrans-Ensemble, which incorporates deep learning models based on both sequence and structural information. The sequence-based model employs the Transformer architecture as base encoders for proteins and aptamers, utilizing pre-trained models ProtBert[21] and RNA-FM[22] for initial embedding of protein and aptamer sequences, respectively. The structure-based model captures primary sequence and secondary structure sequence features of both aptamers and proteins. The outputs of these two models are then concatenated and passed through nonlinear multi-layer perceptron (MLP) layers to produce the final predictions. To address the dataset imbalance, short RNA-protein interaction data was added to increase the number of positive samples. The model achieves reliable prediction accuracy on our aptamer-protein benchmark dataset. Additionally, we employed molecular dynamics simulations to confirm that SelfTrans-Ensemble's sensitivity towards mutations in aptamer sequences. Our approach holds potential to serve as a rapid and reliable screening approach for binding aptamer sequences towards target proteins, improving the cost-effectiveness and efficiency of SELEX in aptamer screening.

## Materials and methods
### Data collection and pre-processing

Aptamer data curation: we constructed aptamer dataset by curating data from four well-established aptamer databases: Aptagen(www.aptagen.com/apta-index/), APTABASE[23], the UTexas Aptamer Database[24], and AptaDB. The dataset curation process is detailed below: 1) Aptagen: Aptagen contributed a total of 790 entries involving aptamer-target interactions, with 432 of these entries corresponding to DNA/RNA aptamer-protein interactions.2) APTABASE: APTABASE provided 605 entries of aptamer-target interactions, including 234 entries pertaining to protein interactions. 3) UTexas Aptamer Database: The UTexas Aptamer Database contains 1495 entries for aptamer-target interactions, out of which 504 entries were specifically identified through a 'Protein' keyword search. We retrieved the corresponding protein sequences by conducting UniProtKB[25] searches based on the best name matches. 4) AptaDB: 635 aptamer-protein interaction entries were obtained by searching with the keyword 'Protein'. To further enrich the studied dataset, the Protein Data Bank (PDB)[26] was also consulted for aptamer-related data using the 'aptamer' keyword. This yielded 216 aptamer-protein complexes. After removing duplicate entries and retaining only those with aptamer lengths less than 200 nt and protein sequence lengths less than 1000 aa, we obtained a total of 1070 positive samples of aptamer-protein pairs containing 991 unique aptamers and 389 unique protein sequences.

RNA data curation: to further augment the available dataset, we utilized the publicly available RNA-Protein database, RPI2825, which is derived from the RNA-Protein complex in the PDB database[27], and the NPInter dataset[28, 29]. Similarly, we selected entries containing RNA sequences with lengths less than 200 nt and protein sequences with lengths less than 1000 aa. The resulting non-redundant dataset contains a total of 864 data points with 444 RNA sequences and 459 protein sequences.

As a result, the final pooled dataset consists of 1422 aptamer/RNA sequences and 848 protein sequences, for a total of 1934 aptamer/RNA-protein interaction entries.

We constituted the negative dataset through both combination and mutation strategies. Specifically, for each aptamer sequence, half of the bases in the sequence were randomly mutated, while the protein sequence remained unchanged. Additionally, we randomly paired aptamers and proteins that were not reported to interact in prior studies, creating negative pairs for our analysis. This process yielded a total of 1934 negative samples.

In summary, our training dataset consists of 1934 positive and 1934 negative samples. Given that a pre-trained large model for RNA was employed, for DNA aptamer sequences, thymine (T) is replaced with uracil (U) in order to convert the sequence to RNA sequences.

**Model**

The SelfTrans-Ensemble model leveraged pre-trained models and the features of the primary sequence and secondary structural sequence, as illustrated in Figure 1, to facilitate the prediction of APIs. The model is comprised of three principal components. 1) Sequence Component: Using pre-trained BERT models (ProtBert and RNA-FM, respectively), biophysical features and contextual information in protein sequences and aptamer sequences are captured for sequence representations. 2) The embedding vector that connects the RNA and protein forms a combined vector. The combined vector is then processed through a Transformer Encoder, which consists of an embedding layer and a positional encoding layer, a multi-head attention layer, and a feed-forward layer, allowing the extraction of high-level features and contextual relationships within the sequence. 3) Following the Transformer Encoder, the output feature vector is then processed through a series of multi-layer perceptron (MLP) layers to produce the sequence-based binding score.

2) Structural Component: 1) The K-mer extraction algorithm is employed to compute the sequence features and structural features of the aptamers and proteins separately, following independent specifications [see **K-mer frequency**]. This process generated combinatorial feature vectors based on the K-mer frequencies of both inputs. 2) The resulting combinatorial feature vectors are then passed through a convolutional neural network(CNN) and a bidirectional long- and short-term memory(BiLSTM) network. The CNN structure is utilized to extract local features of apatemer-protein interactions, while the BiLSTM facilitates the learning of long-range dependencies between these local features across the sequence length. 3) Finally, the feature networks of both the aptamer and protein are combined once again to form a feature vector, which outputs the binding score of the structural information model through the MLP layer.

3) Ensemble binding scores: This process is to map the binding scores obtained from the two models into a fusion space through a linear layer network. Then the output final binding score is restricted to a real number between 0 and 1 using the sigmoid activation function. Sequences with binding scores above the 0.5 threshold are considered to have binding potential. All RNA protein interactions (RPIs) are added to the training set, and the test set consists entirely of APIs.

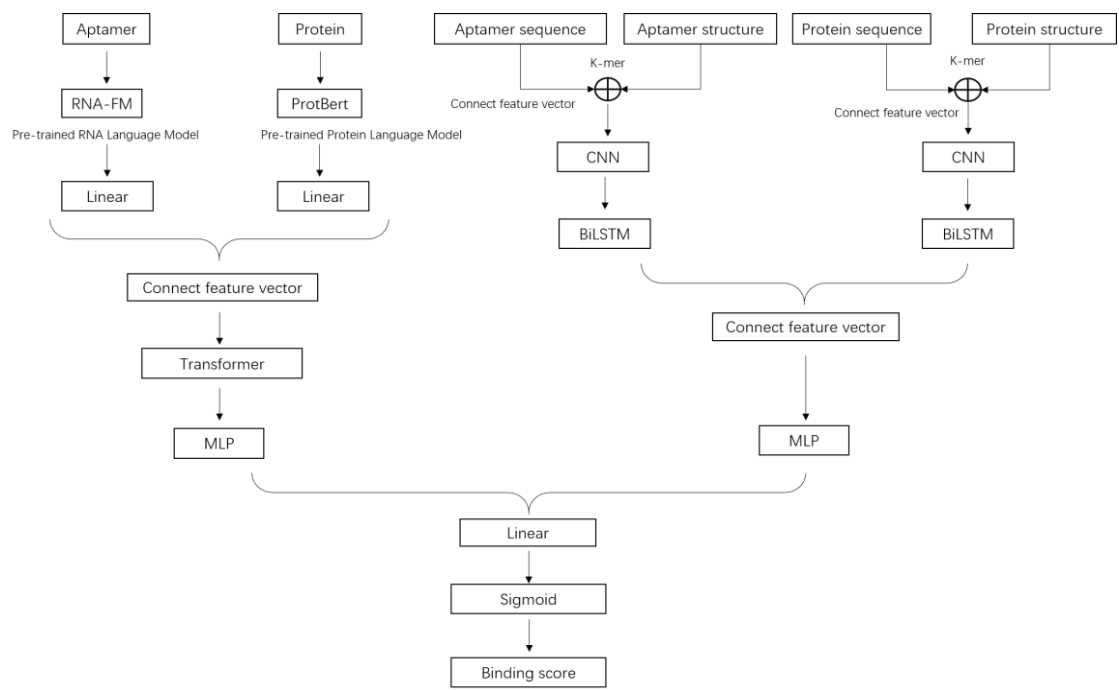

Figure.1 A schematic overview of the training module of SelfTrans-Ensemble.

## Methodology:

### A) Deep learning model for Aptamer-Protein binding prediction

We have developed SelfTrans-Ensemble, a classification model that utilizes the Transformer architecture with pre-trained large models as sequence encoding tools.

**Pre-trained BERT models**

RNA-FM[22] is a pre-trained model based on the BERT architecture, which was obtained by unsupervised training on 23 million unlabeled RNA sequences. It mines the evolutionary and structural patterns embedded in RNA sequences and performs well in many downstream tasks, such as structure prediction, and SARS-CoV-2 variants evolutionary trend prediction[22]. Therefore, we used the RNA-FM model to initialize the coding of RNA sequences. ProtBert[21] is also a pre-trained model rooted in BERT architecture. The ProtBert model is pre-trained using a vast repository of protein sequences in a self-supervised manner, enabling it to capture important biophysical properties of proteins. We harness the ProtBert model to initialize protein sequences. The pre-trained models learned valuable representations from large-scale dataset, making them the effective feature extractor for downstream supervised tasks. Given the constraints imposed by the size of our training dataset, we anticipate that incorporating representations from ProtBert and RNA-FM will enhance both the performance and generalization capabilities of our RNA-protein binding prediction model.

**Transformer Encoder**

Transformer is a seq2seq model proposed in 2017[30]. In this study, we harness the power of the Transformer architecture, specifically the Transformer Encoder. The Transformer Encoder learns the sequential relationships between words and the degree of importance of each word. In this context, we examine the relationship between nucleotide and aptamer sequences, which correspond to words and sentences, respectively, and amino acid and protein sequences as well. This approach extracts features from RNA and protein sequences to serve as the basis for our RNA-protein binding prediction model. We use a three layers Transformer Encoder with two heads for RNA sequences and a three layers Transformer Encoder with eight heads for Protein sequences

**Feature extraction**

Aptanet uses the PseAAC coding method invented by Chou[31] to characterize proteins and k-mer frequency to characterize aptamer or RNA sequences. We integrated their feature extraction approach into the DualBert-Tran-F model as well.

**Pseudo-amino acid composition (PseAAC)**

The PseAAC characterization method is as follows:

A protein chain P consists of N amino acids:

$$P = P_1 P_2 P_3 \cdots P_N \quad (1)$$

The order effect of a protein sequence can consist of a separate set of correlators as follows:

$$\theta_\lambda = \frac{1}{L-\lambda} \sum_{i=1}^{L-\lambda} \Theta(R_i, R_{i+\lambda}) \quad (2)$$

$\theta_\lambda$ are λ-order correlation factors. The function of the correlation factor can be expressed by the following equation:

$$\Theta(R_i, R_j) = \frac{1}{3}\left\{[H_1(R_j)-H_1(R_i)]^2 + [H_2(R_j)-H_2(R_i)]^2 + [M(R_j)-M(R_i)]^2\right\} \quad (3)$$

$H_1(R_i)$, $H_2(R_i)$ and $M(R_i)$ are the values of some physicochemical characteristic of amino acid $R_i$ and similarly $H_1(R_j)$, $H_2(R_j)$ and $M(R_j)$ are the values of some physicochemical characteristic of amino acid $R_j$. Each of the physicochemical characterization values is converted by the following formula:

$$\begin{cases} H_1(i) = \dfrac{H_1^0(i) - \sum_{i=1}^{20} \dfrac{H_1^0(i)}{20}}{\sqrt{\dfrac{\sum_{i=1}^{20} \left[ H_1^0(i) - \sum_{i=1}^{20} \dfrac{H_1^0(i)}{20} \right]^2}{20}}} \\[2em] H_2(i) = \dfrac{H_2^0(i) - \sum_{i=1}^{20} \dfrac{H_2^0(i)}{20}}{\sqrt{\dfrac{\sum_{i=1}^{20} \left[ H_2^0(i) - \sum_{i=1}^{20} \dfrac{H_2^0(i)}{20} \right]^2}{20}}} \\[2em] M(i) = \dfrac{M^0(i) - \sum_{i=1}^{20} \dfrac{M^0(i)}{20}}{\sqrt{\dfrac{\sum_{i=1}^{20} \left[ M^0(i) - \sum_{i=1}^{20} \dfrac{M^0(i)}{20} \right]^2}{20}}} \end{cases} \quad (4)$$

H1(i), H2(i), and M(i) are the initial eigenvalues of the amino acids. Thus, for a protein chain P, the PseAAC can be represented as (20 + λ) dimensions:

$$[V_1, V_2, \cdots V_{20}, V_{21}, \cdots, V_{20+\lambda}]^T \quad (6)$$

T is the transpose symbol.

$$X_u = \begin{cases} \dfrac{f_u}{\sum_{i=1}^{20} f_i + \omega \sum_{j=1}^{\lambda} \theta_j}, & (1 \leq u \leq 20) \\[1em] \dfrac{\omega \theta_{u-20}}{\sum_{i=1}^{20} f_i + \omega \sum_{j=1}^{\lambda} \theta_j}, & (1 \leq u \leq 20) \end{cases} \quad (7)$$

where, for protein chain P, $f_i$ denotes the frequency of occurrence of the 20 amino acids, $\theta_j$ denotes the jth order correlation factor, and ω denotes the weight factor for sequence order effect, here ω is 0.05 and λ is 30. In this study, Aptanet and SelfTrans-Ensemble -F both used 18 physicochemical and biochemical (hydrophobicity, hydrophilicity, side-chain mass, polarity, molecular weight, melting point, transfer free energy, buriability, bulkiness, solvation free energy, relative mutability, residue volume, volume, amino acid distribution, hydration number) properties of amino acids[16].

**K-mer frequency**

K-mer frequency is used in structure-based model and in Aptanet. We converted RNA sequences into DNA sequences for computation when using our own testset on the Aptanet to conform to the Aptanet data input. K-mer is a sliding window mode to represent nucleic acid sequences, and k is the sliding window length. For each bit of sequence in the sliding window, there are four nucleotides(A\T\C\G), so the number of total k-mers is $n^k$. In this study, We have classified the amino acids into seven groups based on dipole moments and side-chain volumes: G1={A,G,V},G2={I,L,F,P},G3={Y,M,T,S},G4={H,N,Q,W}, G5={R,K},G6={D,E},G7={C}. Thus, protein sequences can be represented by the seven letters of AIYHRDC. For protein sequences we use the 3-mer algorithm and the range is extended to 1-3, and for nucleic acid sequences we use the 4-mer algorithm and the range is extended to 1-4. Thus, for protein sequences and protein structural sequences there are $7^1+7^2+7^3$ and $3^1+3^2+3^3$ totalling 438 elements, and for nucleic acid sequences and nucleic acid structural sequences there are $4^1+4^2+4^3+4^4$ and $7^1+7^2+7^3+7^4$ totalling 3140 elements. Aptanet used nucleic acids encoded as k=4 total 256 dimensions.

### SPIDER3 based features

We predicted the secondary structure of proteins using SPIDER3(http://www.sparks-lab.org/server/spider3/), which represents each amino acid in a protein primary sequence in terms of three classical protein secondary structures (α-helix, β-sheet and helix).

### RNAStructure based features

The secondary structures of aptamers and RNAs were predicted by RNAStructure. SPOT-RNA was then used to characterize the secondary structure feature. These structural features are categorized into seven types(letters): S=stem loop, H=hairpin loop, M=multiple loops, I=internal loop, B=bumped loop, X=external loop, and E=terminal. Consequently, the secondary structure sequence of each nucleic acid can be represented with the seven-letter notations.

### Performance evaluation

We validate our trained model by calculating Sensitivity (recall), Specificity, Precision, Accuracy , F1 metrics as follows：

$$\text{Sensitivity} = \frac{TP}{TP+FN} \tag{8}$$

$$\text{Specificity} = \frac{TN}{FP+TN} \tag{9}$$

$$\text{Accuracy} = \frac{TP+TN}{TN+FP+FN+TP} \tag{10}$$

$$\text{Precision} = \frac{TP}{TP+FP} \tag{11}$$

$$F1 = \frac{\text{Precision} \times \text{Sensitivity}}{\text{Precision}+\text{Sensitivity}} \tag{12}$$

### B) Molecular Dynamic for Aptamer-Protein binding prediction

The Amber22 program was used to perform molecular dynamics simulations in this study. High-resolution complex structures were obtained from PDB. For the simulation force field, CHARMM36[32] was chosen, and the solvent water molecules were modeled using the TIP3 model. The nucleic acid molecules and proteins were initially positioned at the center of the water box in the simulation system. The molecules were kept at a distance of 20 Å from the surface of the water box. Additionally, 0.15 mM NaCl and 0.1 mM MgCl2 were added. The simulations were performed using periodic boundary conditions, with the temperature set to 308.15 K, and a simulation time of 500 ns.

### RMSD

RMSD is an important indicator of conformational stability, the smaller the value, the more stable the conformation, then the binding is stable, the formula is as follows:

$$\text{RMSD}(v, w) = \sqrt{\frac{1}{n}\sum_{i=1}^{n}(v_i - w_i)^2} \tag{13}$$

where n denotes the number of atoms, vi is the coordinate vector of the target structure for a given frame, and wi is the coordinate vector of the reference conformation.

### H-bond Frequency

Hydrogen bonding is a crucial intermolecular force for binding in aptamers to proteins. The number of hydrogen bonds formed between the aptamer and the protein directly correlates with the strength of the binding. This study utilized the program Cpptraj H-bond in Amber to count the frequency of hydrogen bond formation between aptamers and proteins.:

$$P = \frac{n}{F} \tag{14}$$

where P denotes the h-bond frequency, n is the number of frames in which hydrogen bonding occurs, and F is the total MD frames.

## Results

### Data distribution

Distribution analysis on the sequence length of aptamers, RNAs and proteins was conducted and shown in Fig.2. Most aptamers are between 40 and 100 nucleotides (nt) in length. Notably, the lengths of RNAs collected to expand the positive data set followed a similar distribution. This intriguing similarity suggests that the interaction between short RNAs and proteins has some similarity to aptamer-protein interactions. On the other hand, the lengths of proteins bound to aptamers were evenly distributed within 600 amino acids, whereas the lengths of proteins bound to RNA were more concentrated within 300 amino acids. This suggests that RNA prefers to bind proteins with shorter sequences.

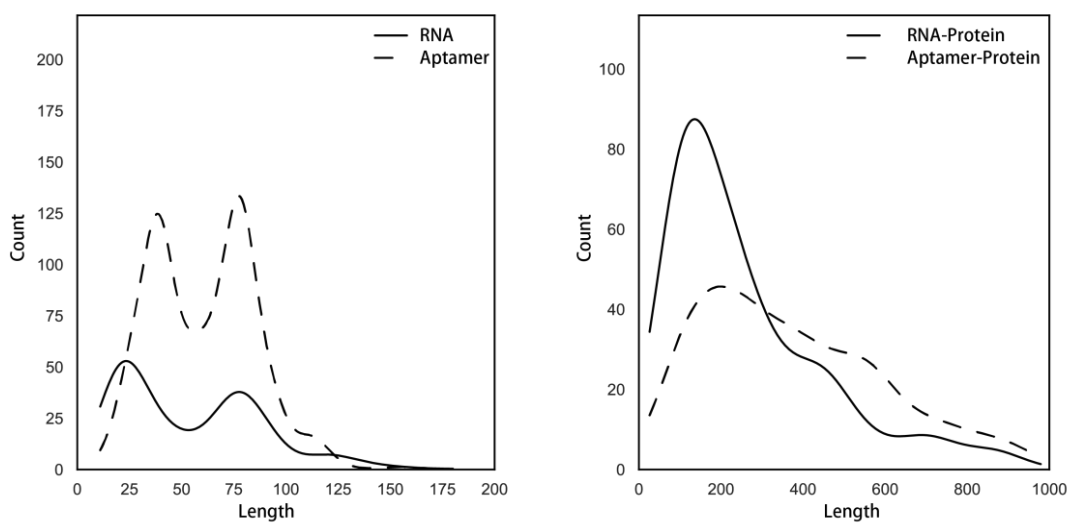

Figure.2 Length distribution of aptamers, RNA, proteins bound to Aptamers and proteins bound to RNA.

### Benchmark study

We compared our DL-based, SelfTrans-Ensemble, approach with Aptanet and AptaTrans in order to further evaluate our performance in terms of model accuracy and generalization ability (Table 1). Aptanet utilizes PseAAC to encode protein features and K-mer to encode aptamer sequence features. we applied them to our own testset. Aptanet, which is the PyTorch encoded version, was trained using our training set and only achieved an F1 metric of 0.624 with our training set and 0.667 with the testset. AptaTrans, on the other hand, was trained from their training set and did not perform well in both our training set and testset, with F1 in both the training set and testset 0.286 and 0.249, respectively. They all perform worse than the SelfTrans-Ensemble model on our testset. Aptanet's poor performance in our dataset may be attributed to the inconsistency between our testset data and the original literature. A possible reason for AptaTrans is that the embedding layer is trained on their own data and may not be as widely applicable as the BERT model.

Table 1 The average performances of SelfTrans-Ensemble compared with Aptanet and AptaTrans

| Model | Dataset | Accuracy | Precision | Recall | F1-score |
|---|---|---|---|---|---|
| Aptanet | Train | 0.624 | 0.625 | 0.992 | 0.767 |
|  | Valid | 0.500 | 0.500 | 1.000 | 0.667 |
| AptaTrans | Train | 0.439 | 0.000 | 0.180 | 0.286 |
|  | Valid | 0.503 | 0.510 | 0.164 | 0.249 |
| SelfTrans-Ensemble | Train | 0.989 | 0.997 | 0.982 | 0.989 |
|  | Valid | 0.880 | 0.854 | 0.943 | **0.896** |

### Ablation study for SelfTrans-Ensemble

We conducted a comprehensive evaluation of different modeling strategies and examined the impact of the sequence-based model and the structural-based model on their predictive performance. Three modeling approaches were employed: 1) DualBert-Trans utilizes sequence information only to predict APIs. 2) Structure-Ensemble, which leverages both primary sequence features and secondary structure sequence features to predict APIs. 3) SelfTrans-Ensemble, which integrates the DualBert-Trans and the Structure-Ensemble models to predict APIs. All models were trained using the same dataset and subsequently tested on a set of independent test sets.

Among these models, the SelfTrans-Ensemble shows a slight advantage over DualBert-Trans on the test set, with an average AUC (Area Under the Curve) score of 0.9232 comparing to 0.9114 AUC score for DualBert-Trans, as shown in Fig.3. The Structure-Ensemble model yields an average AUC score of 0.8559. These results highlight a notable discrepancy in predictive capacity, demonstrating the advantages of language models, particularly in the context of limited training data.

We assess the applicability and generalization capability of the resulting model across varying distributions of positive and negative pairs in order to simulate different real-world scenarios. The F1-score is used as comparison metric as it balances accuracy and recall in the classification model and is considered a distribution-agnostic metric. Other metrics such as accuracy, precision, recall, F1 score are reported in Table 2. The Structure-Ensemble based on structural features decreases the F1 scores with decreasing aptamer proportions. The SelfTrans-Ensemble and the Dualbert-trans converge in F1 scores with different aptamer proportions, and both show significant stability, which illustrates the advantages of language models (Fig.4).

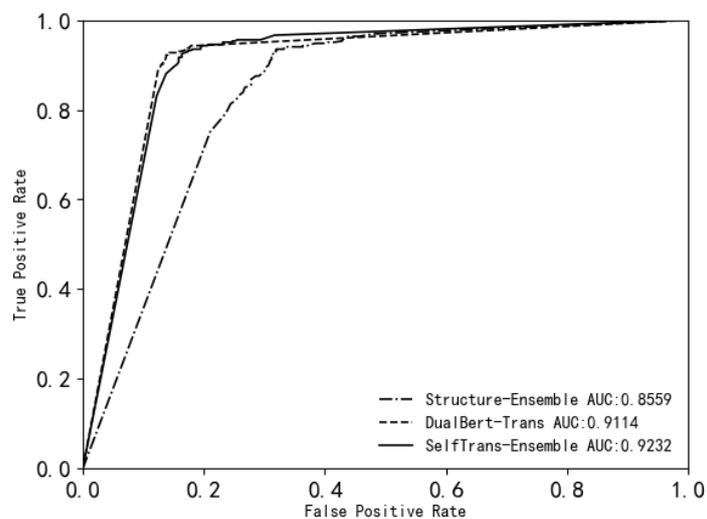

Figure.3 Receiver operating characteristic (ROC) curves of 3 models. Structure-Ensemble, DualBert-Trans, SelfTrans-Ensemble.

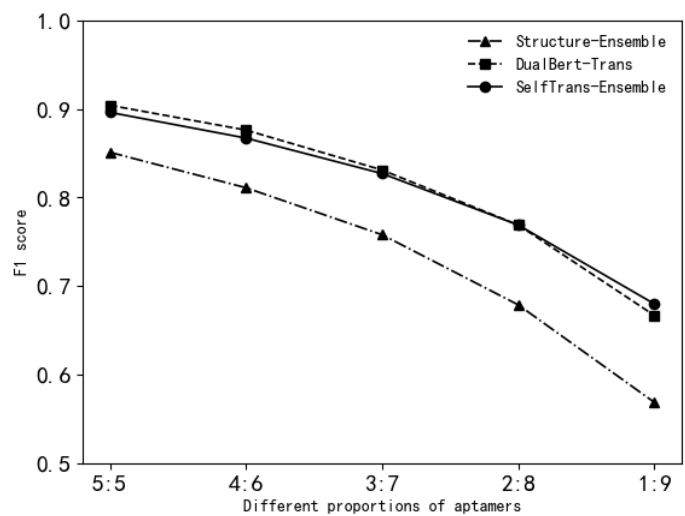

Figure.4 F1-Score variation with aptamer proportion. Structure-Ensemble, DualBert-Trans, SelfTrans-Ensemble.

Table 2 The overall results of our models with our training set and testsets with different proportions of aptamer.

| Model | performance | Valid(5:5) | Valid(4:6) | Valid(3:7) | Valid(2:8) | Valid(1:9) |
|---|---|---|---|---|---|---|
| DualBert-Trans | Accuracy | 0.889 | 0.875 | 0.858 | 0.843 | 0.829 |
|  | Precision | 0.859 | 0.810 | 0.744 | 0.654 | 0.526 |
|  | Recall | 0.953 | 0.953 | 0.941 | 0.931 | 0.912 |
|  | F1-score | 0.904 | 0.876 | 0.831 | 0.769 | 0.667 |
| Structure-Ensemble | Accuracy | 0.820 | 0.799 | 0.778 | 0.753 | 0.731 |

|  | Precision | 0.782 | 0.715 | 0.635 | 0.533 | 0.408 |
|---|---|---|---|---|---|---|
|  | Recall | 0.932 | 0.937 | 0.939 | 0.936 | 0.942 |
|  | F1-score | 0.851 | 0.811 | 0.758 | 0.679 | 0.569 |
| SelfTrans-Ensemble | Accuracy | 0.880 | 0.867 | 0.854 | 0.842 | 0.831 |
|  | Precision | 0.854 | 0.803 | 0.737 | 0.650 | 0.529 |
|  | Recall | 0.943 | 0.943 | 0.941 | 0.843 | 0.953 |
|  | F1-score | 0.896 | 0.867 | 0.827 | 0.769 | 0.680 |

**Attention Layer Analysis**

The design of the attention layer in Transformer neural networks is to capture dependencies between different elements in a sequence, which can be particularly useful in tasks involving sequential or contextual information[33]. In this study, we investigated the attention learned for aptamer and protein sequences to explore the enabling residue/nucleotides for APIs, and evaluate if the applied transformer-based network is capable to capture the short-range and long-range dependencies efficiently for aptamer and protein sequences. We selected an aptamer-protein complex from the PDB database [PDB ID: 3HXO]. This complex consists of a DNA aptamer and Von Willebrand Factor (VWF). The DNA aptamer sequence is as follows: GGCGTGCAGTGCCTTCGGCCGTGCGGTGCCTCCGTCACGC. The secondary structure of this aptamer is comprised of three stem loops, illustrated in Fig.5. However, the three-dimensional structure of the complex indicates that only the stem-loop formed by the nucleotide pair C24-G34 actively engages in the interactions with target protein (Fig.6).

We plot the three aptamer attention layer weights trained by SelfTrans-Ensemble to visually analyze the learned dependencies of different attention layers. The weight of the two bases represented by each grid in the attention layer, where a smaller value is depicted in blue, while a greater value is indicated in red. In the three available attention layers, the weights of the C24-G34 base pairs increased progressively with the increase of the attention layer(Fig.7). This suggests that the attention layer considers this segment of the aptamer to be strongly associated with binding correlation, which is consistent with previous structural analyses.

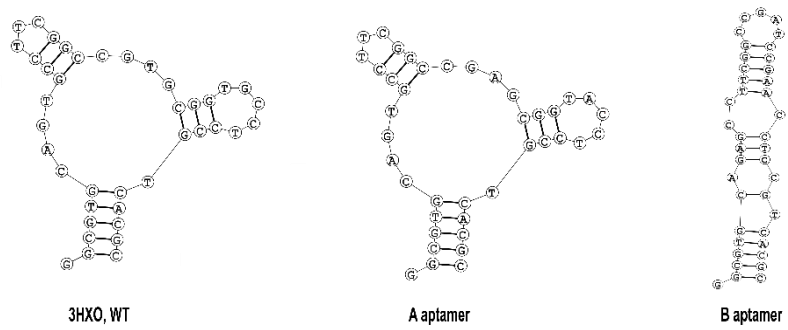

Figure.5 Prediction of secondary structure of 3HXO and its mutants. Secondary structure predicted by RNA Structure[34].

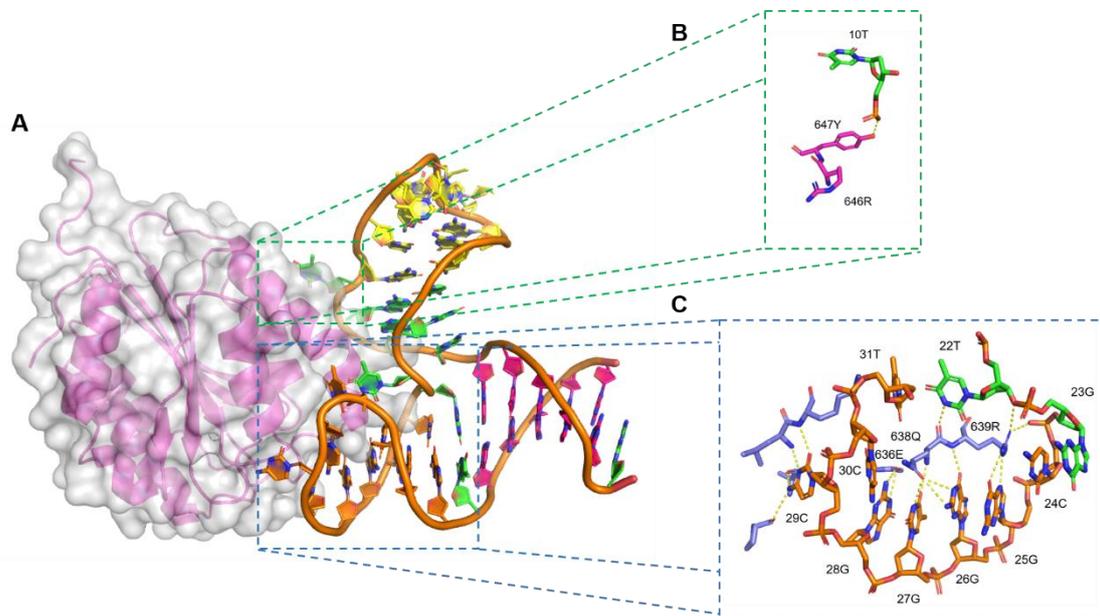

Figure.6 Complex tertiary structure of 3HXO and Hydrogen bonding analysis of the binding site. A) red: 2G-6G with 36C-40C stem, yellow: 11G-19C stem loop, orange: C24-G34 stem loop, violet: protein. B) Hydrogen bonding formed by 10T with 647Y. C) Hydrogen bonding formed by C24-G34 stem loop with 636E, 638Q and 639R.

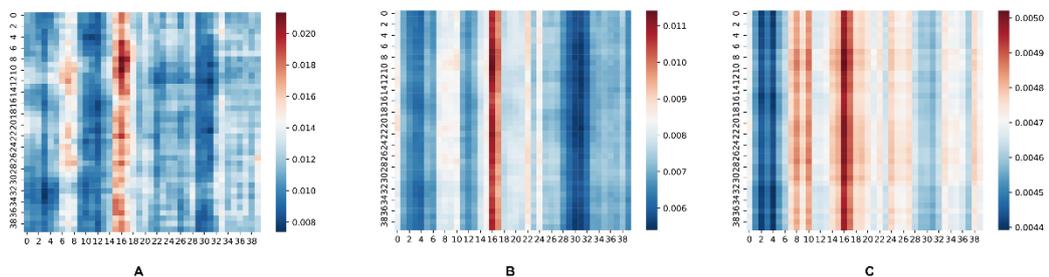

Figure.7 Attention layer weights trained by SelfTrans-Ensemble. A, the first attention layer. B, the second attention layer. C, the third attention layer.

**Molecular dynamic analysis with deep learning-based validations**

In order to assess the model's sensitivity to the aptamer mutations, we envisaged the base mutation of 3HXO. As a result, we conducted a 500 ns molecular simulation to find the hydrogen bonding site between the aptamer and the protein, which will be used as the mutation site. The results of the RMSD analysis indicate that the complex is stably bound, with an RMSD value of 5.18 Å(Fig.8A). Hydrogen bonding analyses (Table 3) reveal that the following hydrogen bonds - 636E-28G, 638Q-22T, 638Q-27T, 639R-23G, 639R-25G, 646R-10T, 647Y-10T (Fig.6) - play a significant role in the binding process.

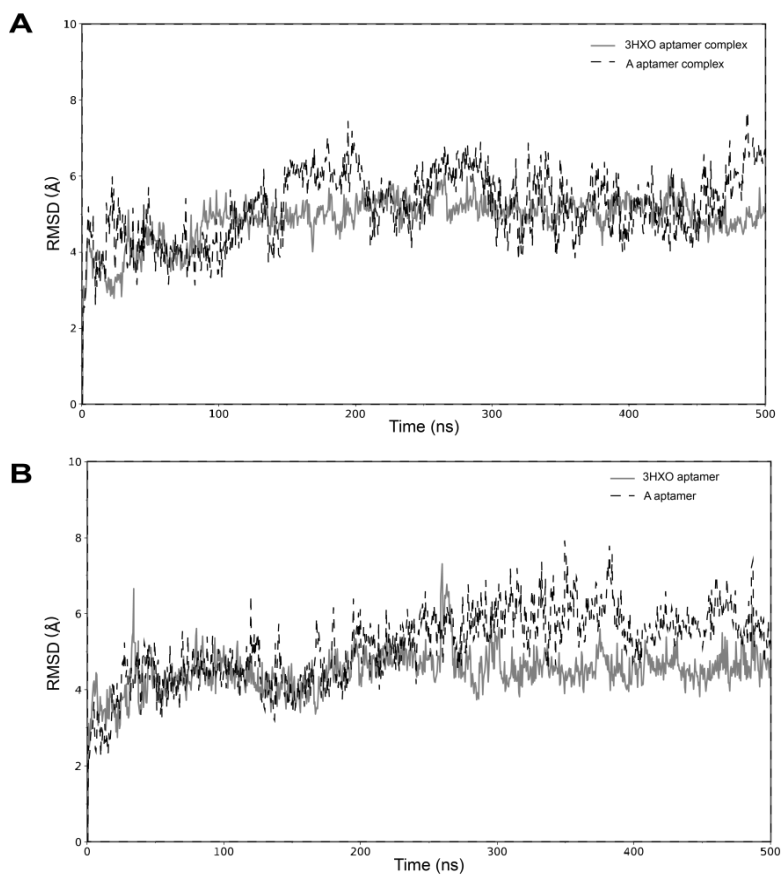

Figure.8 MD RMSDs of the complex(A) and single aptamer(B).

We therefore performed site saturation mutations for 10T, 22T, 23G, 25G, 27T, 28G, and inputted the mutated sequences and proteins into the SelfTrans-Ensemble model.

Here we also predicted all mutated sequences using the DualBert-selfTrans model and the SelfTrans-Ensemble model to assess the model's sensitivity towards mutations. Table 3 shows the number of mutation sites for sequences with a prediction score below 0.5 for each model. When predicting single mutations, SelfTrans-Ensemble assumes that all sequences bind to the target protein, whereas DualBert-SelfTrans assumes that some single mutant sequences do not bind to the target protein. Apparently, the secondary structure of the single mutant sequence is essentially identical to that of the wild type, and this structural match allows the single mutant sequence to maintain binding to the target protein. This illustrates the inadequate sensitivity of the DualBert-SelfTrans model to sequence mutations, despite its comparable performance to SelfTrans-Ensemble in terms of F1 score and AUC. This suggests that the structural modules in the SelfTrans-Ensemble model have a positive effect on the prediction of APIs.

Table 3 number of mutations between SelfTrans-Ensemble and DualBert-selftrans

| Number of mutations | 1 | 2 | 3 | 4 | 5 | 6 | Total |
|---|---|---|---|---|---|---|---|
| number of data | 18 | 135 | 540 | 1215 | 1458 | 729 | 4059 |
| SelfTrans-Ensemble | 0 | 34 | 198 | 604 | 1041 | 611 | 2488 |
| DualBert-Selftrans | 6 | 72 | 351 | 898 | 1160 | 595 | 3082 |

Based on predicted binding scores, aptamer A (T22A, G28A) with a binding score below 0.5 and aptamer B (T10A, T22A, G23T, G25C, T27A, G28A) with the lowest binding score were selected, as illustrated in Table 2. The prediction of the secondary structures of these two sequences (Fig. 5) revealed that the secondary structure of aptamer B had undergone significant structural alterations, resulting in a complete divergence from that of the wild type (WT, 3HXO), whereas the secondary structure of the aptamer A is only partial alterations.

We further conducted molecular dynamics (MD) simulations of 500 ns on the aptamer A to investigate the effect of mutated bases on aptamer-protein binding. The root-mean-square deviation (RMSD) results demonstrated that the binding of aptamer A to the protein exhibited notable fluctuations (Fig. 8A). This can be attributed to the fact that the tertiary structure of aptamer A is less stable than that of the wild-type (WT) type aptamer (Fig. 8B). The analysis of hydrogen bond binding results (Table 5) indicates that the frequency of hydrogen bond formation remained relatively stable for unmutated sites, such as 23G, 25G, and 26G. However, both mutated sites T22A and G28A showed a significant reduction in hydrogen bonding, with no hydrogen bonding formed in some cases. For example, 636E-28G reduced from 0.639 to 636E-28A 0.222, and 132Q did not form hydrogen bonds with the mutated 22A. On the other hand, the mutated 22A formed a new hydrogen bond with 642R. The mutation of the aptamer A results in a reduction or absence of hydrogen bonding between the mutated site bases and the original amino acid. However, the change in tertiary structure also results in the formation of a new hydrogen bond by the aptamer A.

We cross-validated the above result with the SelfTrans-Ensemble model where the threshold set to 0.5. The prediction score decreases from 0.943 for the wild-type (WT) to 0.476 for the mutated aptamer, which aligns with the structural analysis showing fewer hydrogen bonds in the mutated aptamer. While the model predicts that the mutated aptamer no longer binds to the protein, it does so by a narrow margin, as evidenced by the score of 0.476, which is very close to the 0.5 threshold.

Aptanet and AptaTrans were performed on the same case to predict these 2 mutated sequences and the WT sequence (Table 4).

The results demonstrate that the trend of binding scores for Aptanet is analogous to that of the SelfTrans-Ensemble model. This may be attributed to the fact that both employ the K-mer algorithm to extract features from sequences. However, the Aptanet model posits that wild-type aptamer has a reduced probability of binding to the protein. On the other hand, AptaTrans predictions indicate that neither the wild-type nor the mutant sequences display any affinity for the protein. However, AptaTrans also posits that the aptamer A binds to the protein with greater affinity than the B aptamer, which suggests that AptaTrans also captures some information about site mutations. This suggests that the sequence features extracted based on the Transformer model also capture mutation information. The linguistic model of SelfTrans-Ensemble has a Transformer module, while the structural model has a feature extraction algorithm similar to that of Aptanet, which both contributes to improving SelfTrans-Ensemble's sensitivity to mutations.

Table 4 binding scores of 3HXO and mutations

| Name | Sequence | SelfTrans-Ensemble Score | Aptanet Score | AptaTrans Score |
|---|---|---|---|---|
| A | GGCGTGCAGTGCCTTCGGCCGAGCGGTACCTCCGTCACGC | 0.476 | 0.3277 | $1.22 \times 10^{-13}$ |
| B | GGCGTGCAGAGCCTTCGGCCGATCCGAACCTCCGTCACGC | 0.0317 | 0.0632 | $1.85 \times 10^{-15}$ |
| WT | GGCGTGCAGTGCCTTCGGCCGTGCGGTGCCTCCGTCACGC | 0.943 | 0.6494 | $3.16 \times 10^{-11}$ |

Table 5 Hydrogen bonding frequencies of 3HXO and aptamer A in MD

| Amino acids | Nucleotides | WT H-bond Frequency | aptamer A H-bond Frequency |
|---|---|---|---|
| 636E | 28G\28A | 0.639 | 0.222 |
| 638Q | 27T | 0.844 | 0.362 |
| 638Q | 22T | 0.783 | -- |
| 639R | 25G | 0.816 | 0.843 |
| 639R | 23G | 0.600 | 0.547 |
| 639R | 22T | 0.356 | -- |
| 639R | 24C | 0.241 | 0.826 |
| 639R | 26G | 0.375 | 0.336 |
| 642R | 22A | -- | 0.482 |
| 646R | 10T | 0.698 | -- |
| 647Y | 10T | 0.739 | -- |

-- means no hbond

## Discussion

In this study, we introduce the SelfTrans-Ensemble model, a fusion model based on sequence information and structural information. The sequence information-based model employs the Transformer architecture, which has been specifically designed for the processing of RNA and protein sequence data in the field of natural language processing. The module has been designed in a unique manner to facilitate efficient utilisation of pre-trained BERT models. The attention mechanism inherent in the Transformer model is capable of extracting key features from sequence data, thereby assisting in the accurate prediction of aptamer-protein interactions. The structural information-based model employs primary and secondary structural sequence information, combining a convolutional neural network (CNN) with a bidirectional short-term and long-term memory network (BLSTM) to learn features, which facilitates the accurate prediction of aptamer-protein interactions. In the test set, SelfTrans-Ensemble exhibits an F1 score of 0.896 and an AUC of 0.9232, indicating that the model is capable of effectively predicting APIs. We further explored the sensitivity of the model by assessing its response to double mutations in RNA sequences and found that the transformer-based model is capable of capturing small mutations in sequences (Table 2), providing insights of the model's applicability to facilitate RNA design approach aimed at targeting specific proteins. We hypothesized that the enhancement of our model effect is due to the similar binding mechanism with protein shared by aptamer and short RNA. This may help to solve the problem of scarce aptamer data. This also suggests that our model could also predict short RNA-protein interactions.

There is still a lot of room for improvement in this model, and our study only uses the BERT model as a tool for encoding sequences, which can be incorporated into our model to fine-tune the encoding layer to make the BERT model more APIs-adaptive. We are also aware that the current modeling performance is inevitability limited by the size of available training dataset. Therefore, with the availability of better aptamer databases in the future, our model will perform better in APIs prediction. We acknowledge that our negative samples were generated based on the assumption that the mutated aptamer would not bind to the target protein if half of its sequence was mutated, and that an aptamer designed to target one protein would not bind to another protein. Sequence-based models are simple and computationally efficient, but they lack the ability to combine detailed information. This work employs molecular simulation to validate changes in sequence and protein binding hydrogen bonding after mutation. The results demonstrate the model's sensitivity to mutated sequences and highlight the effectiveness of combining molecular simulation with deep learning as a validation method.